\documentclass[
aip,
reprint,
superscriptaddress,
longbibliography,
showkeys,
 amsmath,amssymb,
 aps,
onecolumn
]{revtex4-2}

\usepackage{graphicx}
\usepackage[caption=false]{subfig} 
\usepackage{amsmath}
\usepackage{amssymb}
\usepackage{mathtools}
\usepackage{ bbold }
\usepackage{float}
\usepackage{algorithm}
\usepackage{algpseudocode}
     
\usepackage{dcolumn}
\usepackage{bm}

\newcommand{\ee}{\end{equation}} 
\newcommand{\be}{\begin{equation}}

\usepackage[mathscr]{euscript}  
\usepackage{xcolor}  

\usepackage{tcolorbox}
\usepackage{comment}
\usepackage{float}

\makeatletter
\newsavebox{\@brx}
\newcommand{\llangle}[1][]{\savebox{\@brx}{\(\m@th{#1\langle}\)}%
  \mathopen{\copy\@brx\kern-0.5\wd\@brx\usebox{\@brx}}}
\newcommand{\rrangle}[1][]{\savebox{\@brx}{\(\m@th{#1\rangle}\)}%
  \mathclose{\copy\@brx\kern-0.5\wd\@brx\usebox{\@brx}}}
\makeatother

\begin{document} 

\preprint{APS/123-QED}

\title{Density-dependent stochastic resetting: a large deviations framework for achieving target distributions over networks}






\author{Francesco Coghi}
\thanks{\textit{francesco.coghi@nottingham.ac.uk}}
\affiliation{School of Physics and Astronomy, University of Nottingham, Nottingham, NG7 2RD, United Kingdom}
\affiliation{Centre for the Mathematics and Theoretical Physics of Quantum Non-Equilibrium Systems, University of Nottingham, Nottingham, NG7 2RD, United Kingdom}

\author{Kristian St\o{}levik Olsen}
\thanks{\textit{kristian.olsen@hhu.de}}
\affiliation{Institut für Theoretische Physik II - Weiche Materie, Heinrich-Heine-Universität Düsseldorf, D-40225 Düsseldorf, Germany}

\begin{abstract}

We develop a framework for designing density-dependent stochastic resetting protocols to regulate distributions of random walkers on networks. Resetting mechanisms that depend on local densities induce correlations in otherwise non-interacting walkers. Our framework allows for the study of both transient trajectories and stationary properties and identifies resetting protocols that maximise the likelihood of homogeneous and, more generally, rare configurations of random walkers.
\end{abstract}

\maketitle

\section{Introduction}

Networks arise across numerous scientific disciplines, ranging from social contact networks to transportation and resource distribution systems~\cite{Albert2002,Barrat2008,Newman2010,Latora2017}. When explored through random walks~\cite{Noh2004,Yang2005,Masuda2017,Carletti2020}, these networks naturally reach steady states reflecting their heterogeneous structure---highly connected nodes may attract higher populations, leading to uneven distributions. In many practical scenarios, however, achieving a homogeneous distribution of resources across a network is desirable~\cite{Manfredi2018,Carletti2020,Bassolas2022,DiMeco2024}. For example, in managing utility bicycles in a city, it is crucial to prevent overcrowding at specific locations and ensure their availability at other sites. A warehouse acting as a central stock point could facilitate redistribution by repeatedly resetting utility bicycles from clustered locations to the warehouse, and therefore favour a more balanced configuration across the network. Such strategies are particularly relevant when nodes have finite carrying capacities, where breaches could result in inefficiencies or detrimental effects.

Here, we develop a novel framework to design stochastic resetting protocols that maximise the likelihood of achieving specific configurations, such as homogeneous distributions, despite underlying network heterogeneity. Our resetting mechanism is density-dependent, where the fraction of a population reset from a node depends on its local population. This introduces complex correlations among an otherwise non-interacting population of random walkers. Unlike traditional uses of resetting, which interrupts and re-starts stochastic trajectories\cite{evans2011diffusion,evans2011diffusion2,evans2020stochastic}, our density-dependent approach leverages resetting to counteract certain network-induced dynamics and promote specific global configurations.

\begin{figure}[h!]
    \centering
    \includegraphics[width= 12cm]{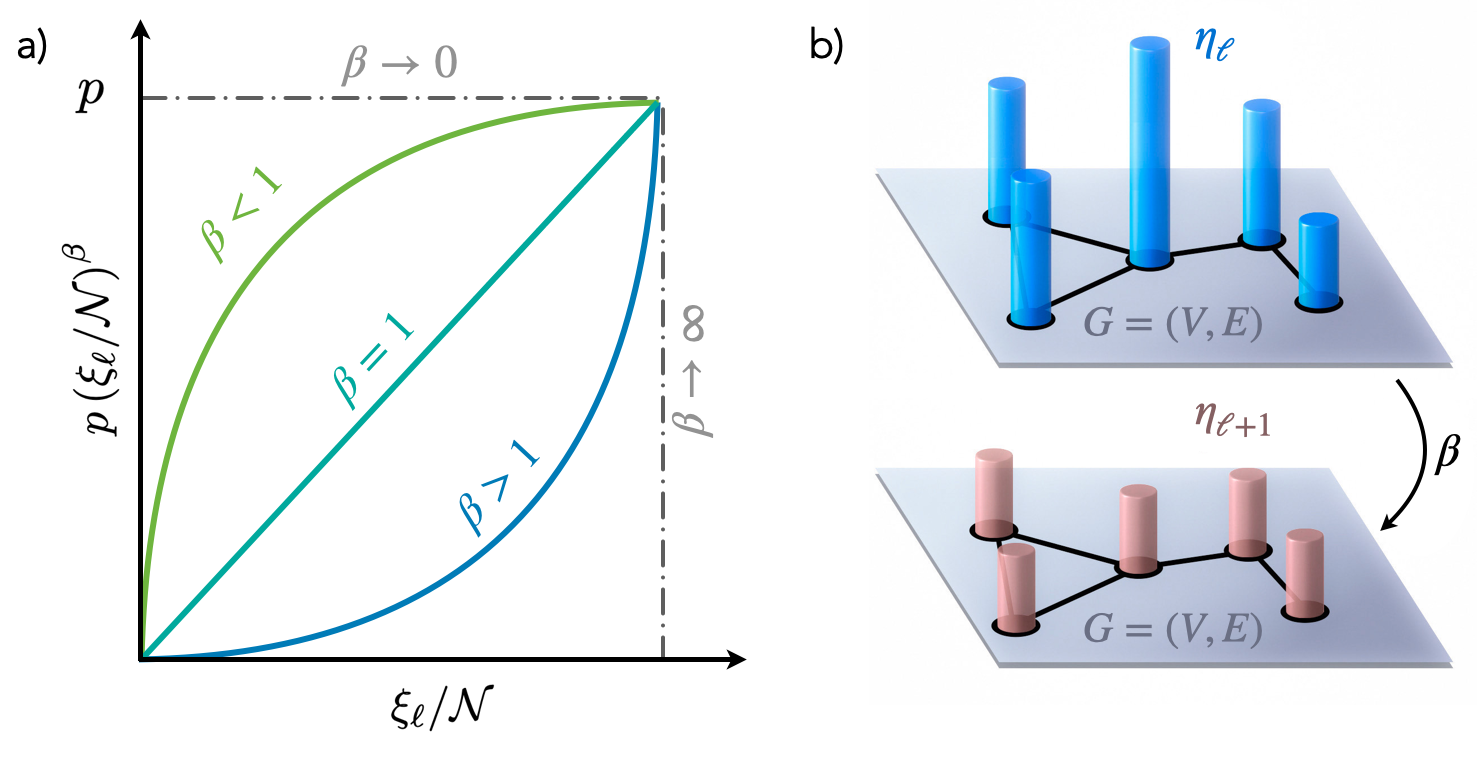}
    \caption{Schematics of the density-dependent resetting protocol. In a), how the probability of resetting $p(\xi_\ell/\mathcal{N})^\beta$ from a node with population $\xi_\ell$ changes at varying $\beta$. In b), a cartoon showing the desired effect of the density-dependent reset scheme at homogenising the walkers population over a graph.}
    \label{fig:densitydep}
\end{figure}

To illustrate this, we study a resetting protocol where the resetting probability at a node scales as a power-law function of the node’s population density (see Fig.\ref{fig:densitydep}). On heterogeneous networks, we find optimal power-law exponents, $\beta$, that maximise the likelihood of achieving homogeneous configurations. Whether this optimal resetting scheme makes homogeneous states typical or just more likely to achieve depends on the network structure.

Our framework is inspired by rare-event sampling techniques, such as genealogy algorithms~\cite{DelMoral2004,DelMoral2005} and cloning algorithms~\cite{Giardina2006,Lecomte2007, Giardina2011,Nemoto2016}, and leverages these methods to sample interesting rare events of the distribution of walkers over graphs. Additionally, our approach extends these techniques by incorporating the resetting mechanism, which potentially introduces an entirely new layer of control. The framework developed here allows for the study of both transient trajectories and long-time stationary properties of node occupation measures on graphs, addressing both dynamical pathways of the system and long-term averaged configurations that are especially relevant in practical applications where full control of the system is not possible.

The paper is organised as follows: in Section \ref{sec:Model}, we introduce the model of density-dependent resetting dynamics and describe both typical and stationary dynamics. Section \ref{sec:Control} focuses on studying fluctuations of the resetting scheme to identify protocols that maximise the likelihood of specific atypical states, such as homogenous state occupancies. Finally, in Section \ref{sec:Examples}, we apply the framework to illustrate how it achieves homogeneous distributions of walkers on a fully connected graph and a heterogeneous graph with non-uniform vertex connectivity. We conclude the paper with a summary and open questions in Section \ref{sec:Conclusion}.


\section{Model of density-dependent reset dynamics}
\label{sec:Model}

\subsection{Model}
\label{ref:Model}

We consider an ensemble of $\mathcal{N}$ discrete-time random walks $\left( X_\ell \right)^{(k)}_{0 \leq \ell \leq n}$, where $k \in [1, \mathcal{N}]$, exploring an undirected, unweighted graph $G = (V, E)$ in $n$ time steps. Here, $V$ denotes a set of $N$ vertices, while $E$ represents the set of edges. At each discrete time step $\ell$, the state $X_\ell$ takes a value in the vertex set $V$. All random walks in the ensemble adhere to the same transition dynamics defined by a transition matrix $\Pi = \left\lbrace \pi_{i,j} \right\rbrace_{i,j \in V}$. Specifically, the entry $\pi_{i,j}$ represents the probability of transitioning from vertex $i$ to vertex $j$ at time $\ell$, given that the random walk was at vertex $i$ at time $\ell-1$.

We further impose that a random walk can move from vertex $i$ to vertex $j$ only if $i$ and $j$ are directly connected in the graph $G$. This connectivity constraint is encoded using the adjacency matrix $A = \left\lbrace a_{i,j} \right\rbrace_{i,j \in V}$, where $a_{i,j} = 1$ if $(i, j) \in E$, and $a_{i,j} = 0$ otherwise.

All $\mathcal{N}$ random walks are initialised at time $\ell = 0$ on a designated vertex $s$, referred to as the \textit{source} node. In our motivating example, this source node functions similarly to a warehouse. At each time step, every random walk independently moves to an adjacent vertex according to the transition probabilities defined by the matrix $\Pi$.

After each step, but before the subsequent time step begins, a fraction of the random walks occupying each vertex is \textit{reset} to the source node. Previously, the effect of resetting on random walks on networks has been studied in the context of steady-states and first-passage times\cite{riascos2020random,chen2022random}, or resetting rules that depend on the network topology \cite{zelenkovski2023random,bowater2023pagerank,huang2021random,ye2022random}. The resetting protocol we consider follows a power-law dependence on the number of walkers present at each occupied vertex. Let $\eta_{\ell+1}(i)$ denote the number of random walks at vertex $i$ at time $\ell + 1$. The update rule for $\eta_{\ell+1}(i)$ is given by:
\begin{equation}
    \label{eq:Dynamics}
    \eta_{\ell+1}(i) = \xi_{\ell+1}(i) - p \left( \frac{\xi_{\ell+1}(i)}{\mathcal{N}} \right)^{\beta} \xi_{\ell+1}(i) + \delta_{i,s} \sum_{j \in V} p \left( \frac{\xi_{N,\ell+1}(j)}{\mathcal{N}} \right)^{\beta} \xi_{\ell+1}(j) \, ,
\end{equation}
where
\begin{equation}
    \label{eq:ProposedMove}
    \xi_{\ell+1}(i) = \sum_{j \in \partial(i)} \eta_{\ell}(j) \pi_{j,i} \, ,
\end{equation}
represents the number of walkers that have moved from a neighbouring node of $i$ to $i$ in the time step $\ell \rightarrow \ell+1$, viz.\ the occupation on node $i$ before resetting takes place. The second term on the r.h.s.\ of $\eqref{eq:Dynamics}$ accounts for the number of walkers removed from vertex $i$ due to the resetting process. These walkers are subsequently placed back at the source node $i = s$, as represented by the third and final term on the right-hand side of \eqref{eq:Dynamics}. The fraction of walkers at vertex $i$ that is reset to the source node is given by
\begin{equation}
    \label{eq:ResetFraction}
    p \left( \frac{\xi_{\ell+1}(i)}{\mathcal{N}} \right)^{\beta} \, ,
\end{equation}
where $p \in [0,1]$ controls the resetting probability, and $\beta \geq 0$ modulates the non-linearity of the reset mechanism. This resetting rule introduces correlations into the dynamics of the random walks, as the probability of resetting depends on the local density of walkers at each vertex.

To better understand the resetting mechanism, let us first examine the special case where $\beta = 0$. In this scenario, the fraction of walkers removed from vertex $i$ at each time step remains constant at $p$, regardless of the number of walkers currently occupying that vertex. This implies that, irrespective of how many random walkers visit vertex $i$ at time $\ell+1$, a fixed proportion $p$ will always be reset to the source node.

In this case, each random walk has an independent probability of being reset after exactly $\tau$ steps, given by $(1-p)^{\tau-1} p$. This probability distribution is geometric with parameter $p$, indicating that the resetting process is memoryless and independent of the movement dynamics of other random walkers. Thus, when $\beta = 0$, the model reduces to a collection of non-interacting and uncorrelated random walkers exploring the graph, with each walker resetting to the source vertex according to an independent geometric distribution. This is analogous to the case of Poissonian resetting in the context of continuous space-time diffusion

In contrast, when $\beta > 0$, the resetting process becomes inherently dependent on the dynamics of all the other walkers. Specifically, the probability $P_\beta(\tau)$ for the $k$-th walker to reset after exactly $\tau$ steps is given by
\begin{equation}
    \label{eq:ResetProbability}
    P_\beta(\tau) = p \left( \frac{\xi_{\tau}\left( X^{(k)}_\tau \right)}{\mathcal{N}} \right)^\beta \, \prod_{\ell=1}^{\tau-1} \left( 1 - p \left( \frac{\xi_{\ell}\left( X^{(k)}_\ell \right)}{\mathcal{N}} \right)^\beta \right) \, .
\end{equation}
Here, $P_\beta(\tau)$ depends on the entire history of occupation counts $\xi_{\ell}(i)$ at each vertex $i$ visited by the $k$-th walker up to time $\tau$. Thus, unlike the $\beta = 0$ case, the resetting probability is no longer independent of the other walkers’ dynamics. Instead, it explicitly incorporates the local density of walkers at the nodes visited along the walker's path. As a result, the dynamics of the random walkers become correlated: while the walkers still do not interact directly at individual time steps, the resetting mechanism indirectly couples their behaviour through the shared occupation statistics. Correlations induced through resetting has also been observed in models with collective simultaneous resetting \cite{Biroli2023,biroli2024dynamically,kulkarni2024dynamically}.

For the purpose of this work, it suffices to recognise that these correlations are induced by the resetting rule. A detailed analysis of these correlations is beyond the scope of this study and will be addressed in future work. For the remainder of this section, we will focus on examining the typical behaviour of the system under the resetting dynamics.

\subsection{Typical dynamics}
\label{ref:TypicalDynamics}

We start by normalising \eqref{eq:Dynamics} by the total number of walkers $\mathcal{N}$ and define the occupation measure vector as
\begin{equation}
    \label{eq:OccupationMeasure}
    \rho_{\ell} =  \frac{\eta_{\ell}}{\mathcal{N}} \, ,
\end{equation}
which represents the density of walkers at each vertex. By taking the limit as $\mathcal{N} \rightarrow \infty$, the normalised occupation measure $\rho_{\ell}$ converges to $\bar{\rho}_\ell$, which takes values in $[0,1]^N$. In this asymptotic limit, we can recast \eqref{eq:Dynamics} to describe the evolution of $\bar{\rho}_\ell$:
\begin{equation}
\label{eq:KineticCorr}
\bar{\rho}_{\ell+1}(i) = \sum_{j \in \partial(i)} \bar{\rho}_{\ell}(j) \pi_{j,i} - p \left( \sum_{j \in \partial(i)} \bar{\rho}_{\ell}(j) \pi_{j,i} \right)^{\beta} \sum_{j \in \partial(i)} \bar{\rho}_{\ell}(j) \pi_{j,i} + \delta_{i,s} \sum_{j \in V} p \left( \sum_{k \in \partial(j)} \bar{\rho}_{\ell}(k) \pi_{k,j} \right)^{\beta} \sum_{k \in \partial(j)} \bar{\rho}_{\ell}(k) \pi_{k,j} \ ,
\end{equation}
where $\partial(i)$ denotes the set of nodes that are nearest neighbors of $i$.

This expression represents the \textit{law of large numbers} (or \textit{typical dynamics}) for an infinitely large ensemble of random walkers exploring the finite graph $G$. It captures the dynamics of the system in the kinetic theory limit, where individual fluctuations are negligible, and the average behaviour dominates.

The fraction of walkers reset to the source node $s$ is given by:
\begin{equation}
    \label{eq:ResetFractionKinetic}
    p \left( \sum_{j \in \partial (i)} \bar{\rho}_\ell(j) \pi_{j,i} \right)^\beta \, ,
\end{equation}
for each vertex $i$ at each time step $\ell$.
When $\beta = 1$, the fraction of walkers being reset is directly proportional to the fraction of walkers visiting vertex $i$ at time step $\ell+1$. Adjusting $\beta$ alters the resetting fraction non-linearly:
\begin{itemize}
    \item For $\beta < 1$, a larger fraction of walkers is reset, implying a higher sensitivity to the presence of walkers at moderately occupied vertices.

    \item For $\beta > 1$, the resetting fraction decreases, reducing the probability of reset, especially at vertices with moderate occupation.
\end{itemize}
Notably, the effect of varying $\beta$ from the case of $\beta = 1$ is more pronounced at vertices with moderate occupation levels, while it becomes weaker at the extremes, such as at highly crowded vertices and sparsely occupied vertices. This behaviour is illustrated graphically in Fig.\ \ref{fig:densitydep}a.

The framework developed here allows us to study fluctuations of $\rho_{\ell}$ around its typical behaviour $\bar{\rho}_\ell$ and, specifically, how to achieve balanced occupation across graph vertices. This is possible because the dynamics in \eqref{eq:KineticCorr} is Markovian, meaning that $\rho_{\ell+1}$ is entirely determined by $\rho_\ell$.

However, the dynamics of an individual random walker, such as $ \left( X_\ell^{(k)} \right)_{0 \leq \ell \leq n}$, is non-Markovian due to the resetting strategy, which makes $X_{\ell+1}^{(k)}$ dependent on $\rho_\ell$. Nevertheless, this does not affect our analysis since we focus on the ensemble dynamics as a whole rather than on individual trajectories. This shift in focus allows us to leverage the Markovian nature of the system at the collective level to understand fluctuations around the typical behaviour.

\subsection{Stationary dynamics}

Before delving into large deviations, we discuss the stationary behaviour of $\bar{\rho}_\ell$ in \eqref{eq:KineticCorr}. We assume that as $\ell \rightarrow \infty$, $\bar{\rho}_\ell$ converges to a stationary distribution $\bar{\rho}_{\text{inv}}$. In such a limit, \eqref{eq:KineticCorr} simplifies to
\begin{equation}
    \label{eq:Stationary}
    \bar{\rho}_{\text{inv}}(i) =  \sum_{j \in \partial(i)} \bar{\rho}_{\text{inv}}(j) \pi_{j,i} \left( 1 - p \left( \sum_{j \in \partial(i)} \bar{\rho}_{\text{inv}}(j) \pi_{j,i} \right)^{\beta} \right) + \delta_{i,s} \sum_{j \in V} p \left( \sum_{k \in \partial(j)} \bar{\rho}_{\text{inv}}(k) \pi_{k,j} \right)^{\beta+1} \, .
\end{equation}
This equation characterises the stationary occupation measure across the graph.
However, it cannot generally be solved analytically. For small graphs, one may attempt numerical solutions. Alternatively, for larger systems, a more practical approach is to directly simulate the long-time evolution of a large ensemble of random walkers according to either \eqref{eq:KineticCorr} or \eqref{eq:Dynamics} with large $\mathcal{N}$ to approximate $\bar{\rho}_{\text{inv}}$. 
The main difficulty in solving \eqref{eq:Stationary} analytically arises from the influence of the parameter $\beta$ on the resetting process, which introduces non-linear dependencies in the stationary distribution. When $\beta=0$ and $p > 0$, \eqref{eq:Stationary} can be explicitly solved for $\bar{\rho}_{\text{inv}}$ as
\begin{equation}
    \label{eq:RhoInvNonCorr}
    \bar{\rho}_{\text{inv}} = p e_s \left( \mathbf{1} - (1-p) \Pi \right)^{-1} \ ,
\end{equation}
where $e_s$ is the canonical Euclidean vector and $\mathbf{1}$ is an $N \times N$-dimensional diagonal matrix of ones. 

In the case of non-interacting and non-correlated random walkers, $\bar{\rho}_{\text{inv}}$ in \eqref{eq:RhoInvNonCorr} can be interpreted as the fraction of time a single random walker spends at each vertex of $G$ in the long-time limit. Due to the resetting mechanism, however, $\bar{\rho}_{\text{inv}}$ does not coincide with the dominant left eigenvector of the original transition matrix $\Pi$.

To account for the impact of resetting, we define an adjusted one-step transition matrix for the process, denoted as
\begin{equation}
    \label{eq:GeneratorNonCorr}
    \Pi^{(R)} = (1-p) \Pi + p \mathbb{1} \, , 
\end{equation}
where $p \mathbb{1}$ represents the reset matrix. Each row of this matrix contains a value of $p$ in the column corresponding to the source node $s$, with zeros elsewhere. This construction reflects the probability of either transitioning according to $\Pi$ (with probability $1-p$) or resetting to the source node $s$ (with probability $p$). It can be shown that $\bar{\rho}_{\text{inv}}$ in \eqref{eq:RhoInvNonCorr} is the dominant left-eigenvector of $\Pi^{(R)}$ in \eqref{eq:GeneratorNonCorr}. 

In contrast, when $p > 0$ and $\beta > 0$, a Markov generator of the form \eqref{eq:GeneratorNonCorr} no longer exists because, although the random walkers do not interact directly, they become correlated through the resetting mechanism. This interdependence means that $\bar{\rho}_{\text{inv}}$ in \eqref{eq:Stationary} cannot, in general, be expressed in a simple closed form. 

While, in principle, one could construct a Markov generator for the entire ensemble of walkers, rather than for each walker individually, this would require considering the joint evolution of all walkers. Since the state of the ensemble at the next time step depends on the collective occupation across the graph, the state space of such a Markov process would be extremely high-dimensional (on the order of $e^{\mathcal{N}}$ rather than just $N$). As a result, analysing its spectrum would necessitate alternative methods, which fall outside the scope of this work. Therefore, we will not explore this direction further here.

\subsection{Examples}

\subsubsection{Fully-connected graph (FCG) example}

Here, briefly, we derive the typical behaviour of \eqref{eq:KineticCorr} as well as the stationary distribution \eqref{eq:Stationary} in the simple case of a fully-connected graph of $N$ vertices without self-loops. In such a case, the transition matrix for the random walkers is given by
\begin{equation}
    \label{eq:TransitionMatrixFully}
    \pi_{ij} = \frac{(1-\delta_{ij})}{N-1} \, .
\end{equation}

In the simplest scenario of $\beta=0$, \eqref{eq:KineticCorr} simplifies to
\begin{equation}
\label{eq:KineticFully} 
    \begin{cases}
    \bar{\rho}^*_{\ell+1} &= \left( \frac{N-2}{N-1} \bar{\rho}_{\ell}^* + \frac{1}{N-1} \bar{\rho}_\ell(s) \right) (1-p) \\
    \bar{\rho}_{\ell+1}(s) &= 1 - (N-1)\bar{\rho}_{\ell+1}^*  \, ,
\end{cases}
\end{equation}
having that all vertices, but the source, are statistically equivalent in the typical scenario, i.e, $\bar{\rho}^*_\ell \equiv \bar{\rho}^*_\ell(i)$ for all $i \neq s$. At stationarity, considering eq.\ \eqref{eq:Stationary} for $\beta=0$, or \eqref{eq:KineticFully} with $\bar{\rho}_\ell = \bar{\rho}_{\ell+1} \eqqcolon \bar{\rho}_{\text{inv}}$, we obtain
\begin{equation}
\label{eq:StationaryFully}
    \begin{cases} 
    \bar{\rho}_{\text{inv}} &= \frac{1-p}{N - p} \\
    \bar{\rho}_{\text{inv}}(s) &= \frac{1 + (N-2) p}{N - p} \ .
    \end{cases}
\end{equation}

When $\beta>0$, similarly to \eqref{eq:KineticFully}, \eqref{eq:KineticCorr} becomes
\begin{equation}
\label{eq:KineticFullyBeta} 
    \begin{cases}
    \bar{\rho}^*_{\ell+1} &= \left( \frac{N-2}{N-1} \bar{\rho}_{\ell}^* + \frac{1}{N-1} \bar{\rho}_\ell(s) \right) \left( 1-p \left( \frac{N-2}{N-1} \bar{\rho}_{\ell}^* + \frac{1}{N-1} \bar{\rho}_\ell(s) \right)^\beta \right) \\
    \bar{\rho}_{\ell+1}(s) &= 1-(N-1)\bar{\rho}^*_{\ell+1} \, .
\end{cases}
\end{equation}

At stationarity, we can derive an asymptotic expansion for $\bar{\rho}_{\text{inv}}$ as follows. We expand \eqref{eq:Stationary} for $N$, the number of vertices in the graph, for source $s$ and any other node $i$ as the system
\begin{equation}
    \label{eq:SysStatFully}
    \begin{cases}
    \bar{\rho}_{\text{inv}}(s) &= \bar{\rho}_{\text{inv}}^* \left( 1 - p \left( \bar{\rho}_{\text{inv}}^* \right)^{\beta} \right) + p \left( \bar{\rho}_{\text{inv}}^* \right)^{\beta+1} + p (N-1) \left( \frac{\bar{\rho}_{\text{inv}}(s)+(N-2)\bar{\rho}^*_{\text{inv}}}{N-1} \right)^{\beta+1} \\
    \bar{\rho}^*_{\text{inv}} &= \frac{1-\bar{\rho}_{\text{inv}}(s)}{N-1} \, .
    \end{cases}
\end{equation}
Replacing the second equation in the first one we obtain
\begin{equation}
    \label{eq:SysStatSourceFully}
    N \bar{\rho}_{\text{inv}}(s) = 1 + \frac{p}{(N-1)^{\beta-1}} \left( \bar{\rho}_{\text{inv}}(s) + \frac{N-2}{N-1} (1-\bar{\rho}_{\text{inv}}(s)) \right)^{\beta+1} \, ,
\end{equation}
which can be solved expanding asymptotically for large $N$ by
\begin{equation}
        \label{eq:SysStatSourceFullyPertExp}
        \bar{\rho}_{\text{inv}}(s)= N^{-1} + p N^{-\beta} - p (1+\beta) N^{-1-\beta} + O(N^{-2-\beta}) \, .
\end{equation}

From this we see that the zero-th order term corresponds to the case without resetting, where the distibution is uniform. The effect of resetting enters through the correction terms where the occupation is determined by both the resetting protocol through $p$ and $\beta$.

\subsubsection{Heterogeneous random graphs}
\label{sec:BA}

\begin{figure}
    \centering
    \includegraphics[width=0.7\linewidth]{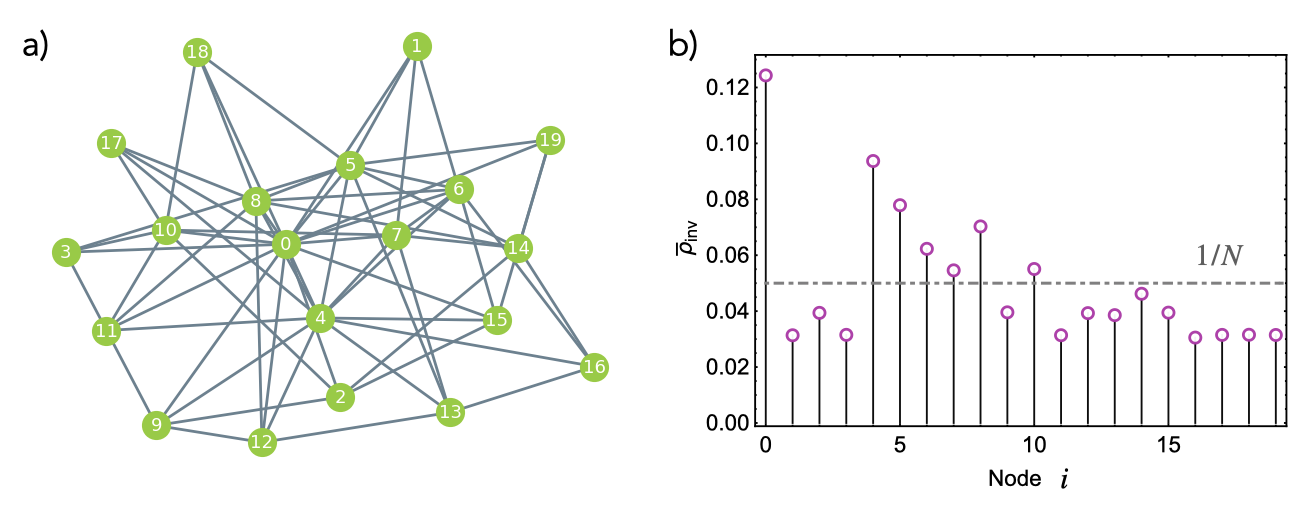}
    \caption{In a) an exemplary BA graph of 20 nodes realised with the algorithm discussed in the text by setting $m=4$. Node $0$ is the most connected and is identified as a hub of the graph. In b) we plot the stationary distribution $\bar{\rho}_{\text{inv}}$ of a large number of non-interacting random walkers exploring the graph. Evidently, such a distribution is not uniform over the graph and walkers spend most of their time on node $0$.}
    \label{fig:BA_typical}
\end{figure}

FCGs provide a useful framework for developing intuition, as they often allow for analytical approximations, as shown above. However, real-world environments are rarely well-approximated by fully connected structures. For this reason, we introduce a different graph structure that we will use in subsequent analyses, as well as illustrate how the stationary distribution $\bar{\rho}_{\text{inv}}$ from \eqref{eq:Stationary} behaves for this structure as the parameters $p$ and $\beta$ vary.

While the graph model we introduce is still far from capturing the complexity of real-world networks, our primary aim here is not to match empirical data. Instead, we seek to explore how heterogeneity in graph connectivity can lead to phenomena such as vertex overcrowding. This simpler, yet structured, graph model is therefore well-suited to our purpose, allowing us to show how variations in the resetting dynamics can impact the distribution of walkers across nodes.


We will consider the Barabási–Albert (BA) model~\cite{Albert2002}, which generates networks that grow over time based on the principle of preferential attachment. This model captures the ``rich get richer" phenomenon observed in many real-world networks, such as social networks, where highly connected nodes are more likely to attract additional connections.

The construction of a BA network proceeds as follows:
\begin{enumerate}
    \item Initialisation: Start with a small, fully connected seed network consisting of $m_0$ vertices.
    \item Growth and Attachment: For each new vertex added to the network, establish $m$ edges between the new vertex and existing vertices. The probability of connecting the new vertex to an existing vertex $i$ is proportional to the degree of vertex $i$, i.e., the number of edges connected to $i$.
\end{enumerate}
As the network grows, this process results in a scale-free network, characterised by a degree (or connectivity) distribution that follows a power law. In such networks, a small number of nodes, known as hubs, accumulate a disproportionately large number of connections, while the majority of nodes have relatively few.

BA random graphs are inherently heterogeneous, with a small number of highly connected hubs and many sparsely connected nodes. This heterogeneity can promote the clustering of walkers in regions with high connectivity, such as hubs.

In Fig.\ \ref{fig:BA_typical} we show an exemplary BA graph as well as the solution of \eqref{eq:Stationary} for $p=0$ for such a structure, which highlights how the heterogeneity in the connections of the graph typically leads to higher and lower occupation nodes. The node with highest occupation is $0$, which we identify as a hub.

In the following, we use large deviation theory to develop an algorithm that allows us to control over the distribution of walkers across any graph, both at each time step and in the stationary state. This approach can be used, for example, to generate homogeneous distributions over the graph, thereby preventing overcrowding in specific regions.

\section{Controlling state occupancy and congestion via large deviation theory}
\label{sec:Control}

In this section, we use large deviation theory to characterise the likelihood of atypical population distributions and to design resetting protocols for controlling these distributions over time. Large deviations have already been applied to stochastic resetting systems, particularly to single-particle systems and most often in the context of long-time large deviations~\cite{Meylahn2015,Harris2017,Zamparo2019a,Coghi2020,Zamparo2021,Zamparo2022}. In contrast, our approach here is different as we consider the number of particles as the scaling variable. We begin by examining trajectorial control, focusing on guiding the system along specific occupation measure trajectories at all times. We then transition to the long-time limit, studying protocols that maximise the likelihood of achieving particular stationary occupation measures, which represent the system’s time-averaged behaviour.

\subsection{All-time control via sample-path large deviations}

We focus on the stochastic occupation measure defined in \eqref{eq:OccupationMeasure}, which is reported explicitly below for clarity, 
\begin{equation}
    \label{eq:OccupationMeasureExplicit}
    \rho_\ell = \frac{1}{\mathcal{N}} \sum_{k=1}^{\mathcal{N}} \delta_{X^{(k)}_{\ell}} \, ,
\end{equation}
and aim at studying the fluctuations around the solution $\bar{\rho}_\ell$ of the kinetic equation \eqref{eq:KineticCorr}. 

We assume that the sample-path Large Deviation Principle (LDP)
\begin{equation}
\label{eq:PathLDPCorr}
\mathbb{P} \left( \left\lbrace \phi_\ell \right\rbrace_{0 \leq \ell \leq n} \right) = e^{- \mathcal{N} \mathcal{I}_{[0,n]} \left(\phi_0, \phi_1, \cdots, \phi_\ell, \cdots \phi_n \right) + o(\mathcal{N})} \ ,
\end{equation}
holds for the probability distribution $$ \mathbb{P} \left( \left\lbrace \phi_\ell \right\rbrace_{0 \leq \ell \leq n} \right) \coloneqq \mathbb{P}\left( \left\lbrace \rho_{\ell} \right\rbrace_{0 \leq \ell \leq n} = \left\lbrace \phi_\ell \right\rbrace_{0 \leq \ell \leq n} \right) \, ,$$
and calculate the large deviation rate function $\mathcal{I}_{[0,n]}$, which characterises the rate of decay of fluctuations away of $\left\lbrace \bar{\rho}_\ell \right\rbrace_{0 \leq \ell \leq n}$ for large $\mathcal{N}$. We point the reader to~\cite{Touchette2009,Grafke2019,Gabrielli2020} for details on large deviations and, in particular, on sample-path large deviations. Given the chosen reset dynamics, not all possible vectors $\phi_\ell$ can be considered valid outcomes of the occupation measure $\rho_\ell$. Firstly, we must require that $0 \leq \phi_\ell(i) \leq 1$ for all $i \in V$ as well as $\ell \in [0, \cdots, n]$, and the usual normalisation condition $\sum_{i \in V} \phi_\ell(i) = 1$. Another, perhaps less obvious, constraint to point out is given by
\begin{equation}
    \label{eq:ConstraintMinimal}
    \phi_\ell(s) \geq p \left( \frac{1}{N-1} \right)^\beta \, ,
\end{equation}
where the r.h.s.\ represents the extreme case where the smallest fraction of walkers is reset to the source node. This occurs under the assumption of a uniform distribution of walkers across all vertices, with the source node $s$ itself being empty. This scenario sets a lower bound on the fraction of walkers that we must expect to find at the source node $s$. Additional constraints will emerge as we further develop the framework, and we will collect and discuss these in a dedicated subsection below.

Considering that the system of resetting random walkers is Markov when we extend the state space to include the occupation measure, we can express the rate function as follows:
\begin{equation}
\label{eq:RateFunctionalCorr}
\begin{split}
\mathcal{I}_{[0,n]} \left( \phi_0, \phi_1, \cdots, \phi_\ell, \cdots \phi_n \right) = \mathcal{I}_0 \left( \phi_0 \right) + \sum_{\ell=1}^n \mathcal{I}_{\ell-1,\ell} \left( \phi_{\ell-1}, \phi_\ell \right) \ ,
\end{split}
\end{equation}
where the factorisation of the probability $\mathbb{P} \left( \left\lbrace \phi_\ell \right\rbrace_{0 \leq \ell \leq n} \right)$ translates into a summation at the level of the rate function $\mathcal{I}_{[0,n]}$. This formulation expresses $\mathcal{I}_{[0,n]}$ in terms of two-component rate functions, $\mathcal{I}_{\ell-1,\ell} \left( \phi_{\ell-1}, \phi_\ell \right)$, which capture the transition costs between consecutive time steps.

Furthermore, since we consider the deterministic initial condition $\rho_0 = \delta_s$, where all random walkers are initially located at the source node $s$ at time $0$, there is no stochasticity associated with this initial state. Thus, we set $\mathcal{I}_0(\phi_0) = 0$.

By introducing the pair occupation measure matrix, similarly to~\cite{Carugno2022},
\begin{equation}
    \label{eq:PairOccupationMeasure}
    L_{\ell} = \frac{1}{\mathcal{N}} \sum_{k = 1}^\mathcal{N} \delta_{X^{(k)}_{\ell-1}} \delta_{X^{(k)}_\ell} \, ,
\end{equation}
where the entry $(i,j)$ represents the fraction of random walkers moving from vertex $i$ to vertex $j$ in one time step $\ell-1 \rightarrow \ell$ (before potentially resetting to the source node), we can express the rate function via the following variational formula for the single two-component rate function $\mathcal{I}_{\ell-1,\ell}$:
\begin{equation}
\label{eq:RateFunctionalVariationalCorr}
\mathcal{I}_{\ell-1,\ell} \left( \phi_{\ell-1} , \phi_\ell \right) = \inf_{\substack{L_\ell \\ \sum_{j \in \partial(i)} L_\ell(i,j) = \phi_{\ell-1}(i) \\ \Omega_\ell(i) \sum_{j \in \partial(i)} L_\ell(j,i) =  \phi_\ell(i)}} \mathcal{H} \left( L_\ell | \phi_{\ell-1} \right) \ ,
\end{equation}
where $\mathcal{H} (L_\ell | \phi_{\ell-1})$ is a non-negative, higher-order rate function that encapsulates the stochastic dynamics governing the one-step evolution of the ensemble of random walkers. It reads
\begin{equation}
    \label{eq:RateFunctionHigher}
    \mathcal{H} (L_\ell| \phi_{\ell-1}) = \sum_{i,j \in V} L_\ell(i,j) \ln \left( \frac{L_\ell(i,j)}{\pi_{i,j} \phi_{\ell-1}(i)} \right) \, ,
\end{equation}
with its derivation provided in Appendix \ref{Appendix:RateFunctionHigher} for the sake of brevity. 

In the variational problem in \eqref{eq:RateFunctionalVariationalCorr}, there are two constraints that govern the dynamics of walkers transitioning from $\phi_{\ell-1}$ to $\phi_\ell$:
\begin{itemize}
    \item The first constraint, $\sum_{j \in \partial(i)} L_\ell(i, j) = \phi_{\ell-1}(i)$, ensures that the total fraction of walkers making a one-step transition from vertex $i$ at time $\ell-1$ (each according to $\Pi$) matches the fraction of walkers initially present at $i$ at time $\ell-1$.

    \item The second constraint, $\Omega_\ell(i) \sum_{j \in \partial(i)} L_\ell(j, i) = \phi_\ell(i)$, is better understood in reverse: given the occupation measure $\phi_\ell(i)$ at time $\ell$, this fraction must correspond to walkers that arrived at $i$ in the previous step $\ell \leftarrow \ell-1$. Here, $\Omega_\ell(i)$ re-weights the fraction to account for the possibility that some walkers may have been reset to the source node before being counted at $i$.
\end{itemize}

These two constraints define the two stages of the density-dependent resetting scheme: free evolution and resetting. These stages are analogous to the steps in genealogy algorithms~\cite{DelMoral2004,DelMoral2005} or cloning algorithms~\cite{Giardina2006,Lecomte2007,Giardina2011,Nemoto2016} used to sample rare events. In these algorithms, the process alternates between free evolution and filtering, with the latter typically involving the selective removal (killing) and replication (cloning) of particles that are most likely to achieve the rare event of interest.

The term $\Omega_\ell(i)$ appearing in \eqref{eq:RateFunctionalVariationalCorr} explicitly reads 
\begin{equation}
\label{eq:OmegaCorr}
\begin{split}
\Omega_{\ell}(i) &= \left( 1 - p \left( \sum_{j \in V} L_\ell(j,i) \right)^\beta \right) + \delta_{i,s} \frac{\sum_{m \in V} p \left( \sum_{j \in V} L_\ell(j,m) \right)^{\beta+1} }{\sum_{j \in V} L_\ell(j,i) } \\ 
&= \left( 1 - p \sigma_\ell^\beta(i) \right) + \delta_{i,s} \frac{\sum_{m \in V} p  \sigma_\ell^{\beta+1}(m) }{\sigma_\ell(i)} \ ,
\end{split}
\end{equation}
where, for convenience, we denote
\begin{equation}
    \label{eq:VirtualDensity}
    \sigma_\ell(i) = \sum_{j \in V} L_\ell(j,i) \, ,
\end{equation}
which represents the fraction of walkers transitioning from any vertex $j$ to vertex $i$ at time step $\ell$. From \eqref{eq:OmegaCorr}, $\Omega_\ell(i)$ captures how many walkers will remain at vertex $i$ after taking into account the resetting mechanism:
\begin{itemize}
    \item For non-source nodes ($i \neq s$): A fraction $p \sigma_\ell^\beta(i)$ of walkers is reset to the source node $s$, where $\sigma_\ell(i)$ is the fraction of walkers arriving at $i$. The term $\left( 1 - p \sigma_\ell^\beta(i) \right)$ then represents the fraction of walkers that stay at $i$.

    \item For the source node ($i = s$): An additional term is included to account for all walkers reset from other nodes back to the source. This term depends on the total resetting contributions from other vertices and adjusts for the density $\sigma_\ell(i)$ at the source.
\end{itemize}

The variational problem in \eqref{eq:RateFunctionalVariationalCorr} can be simplified by introducing the Lagrangian 
\begin{equation}
\label{eq:LagrangianForm}
\begin{split}
\mathcal{L} &= \mathcal{H} (L_\ell| \phi_{\ell-1}) -  \sum_{i \in V} \kappa_\ell(i) \left( \phi_{\ell-1}(i) - \sum_{j \in V} L_\ell(i,j) \right) + \\
&\hspace{1cm} -  \sum_{i \in V} \eta_\ell(i) \left( \phi_\ell(i) - \Omega_\ell(i) \sigma_\ell(i) \right) - \sum_{i \in V} \xi_\ell(i) \left( \sigma_\ell(i) - \sum_{j \in V} L_\ell(j,i) \right) \ ,
\end{split}
\end{equation}
where $\kappa_\ell$, $\eta_\ell$, and $\xi_\ell$ are vectors representing Lagrangian multipliers fixing the respective constraints. We also remark that the last constraint in \eqref{eq:LagrangianForm} is not necessary, but it could be helpful in decoupling the Euler--Lagrange equations associated to the critical points of $\mathcal{L}$. These are $N(N+1)$ equations and read
\begin{equation}
\label{eq:EulerLagrangeCorr}
\begin{cases}
0 &= \frac{\partial \mathcal{L}}{\partial L(i,j)} = \ln \left( \frac{L(i,j)}{\pi_{i,j} \phi_{\ell-1}(i)} \right) + 1 + \kappa_\ell(i) + \xi_\ell(j) \\
0 &= \frac{\partial \mathcal{L}}{\partial \sigma_\ell(i)} = \eta_\ell(i) \Omega_\ell(i) + \sum_{j \in V} \sigma_\ell(j) \eta_\ell(j) \frac{\partial \Omega_\ell(j)}{\partial \sigma_\ell(j)}  - \xi_\ell(i) \ ,
\end{cases}
\end{equation}
with
\begin{equation}
\label{eq:OmegaDerivativeCorr}
\frac{\partial  \Omega_\ell(h)}{\partial \sigma_\ell(j)} = \delta_{h,j} (- p \beta \sigma_\ell^{\beta-1}(j)) + \delta_{h,s} p \left( \frac{(\beta+1)\sigma_\ell^\beta(j)}{\sigma_\ell(h)} - \delta_{h,j} \frac{\sum_{m \in V} \sigma_\ell^{\beta+1}(m)}{\sigma_\ell^2(m)} \right) \, .
\end{equation}

Solving the first equation in \eqref{eq:EulerLagrangeCorr} for $L(i,j)$ gives
\begin{equation}
\label{eq:PairOccupationExplicit}
L(i,j) = \pi_{ij} \phi_{\ell-1}(i) u_\ell(i) v_\ell(j) \, ,
\end{equation}
where we denoted $u_\ell(i) = e^{-\kappa_\ell(i) - 1}$ and $v(j) = e^{-\xi_\ell(j)}$. 

By considering $i \neq s$ and applying the three constraints (i) $\phi_{\ell-1}(i) = \sum_{j \in V} L(i,j)$, (ii) $\sigma_\ell(i) = \sum_{j \in V} L(j,i)$, and (iii) $\phi_\ell(i) = \Omega_\ell(i) \sigma_\ell(i)$ with $\Omega_\ell(i)$ in \eqref{eq:OmegaCorr} we obtain the following set of equations
\begin{eqnarray}
\label{eq:EqStochasticityCorr}
u_\ell(i) &=& \frac{1}{\sum_{j \in V} \pi_{ij} v_\ell(j)} \\
\label{eq:EqForwardChapmanKolmogorovCorr}
v_\ell(j) &=& \frac{\sigma_\ell(j)}{\sum_{i \in V} \pi_{ij} \phi_{\ell-1}(i) u_\ell(i)} \\
\label{eq:ConstraintRho3Corr}
\phi_\ell (j) &=& \sigma_\ell(j) \left( 1-p \sigma_\ell^\beta(j) \right) \ ,
\end{eqnarray}
which need to be solved for $u_\ell$, $v_\ell$, and $\sigma_\ell$ (details will be provided in a dedicated subsection below). When $i=s$, equations \eqref{eq:EqStochasticityCorr} and \eqref{eq:EqForwardChapmanKolmogorovCorr} remain valid, whereas the third equation must be replaced with the normalisation condition
\begin{equation}
\label{eq:NormalisationSigma}
\sigma_\ell(s) = 1 - \sum_{i \neq s} \sigma_\ell(i) \, .
\end{equation}

Given the explicit form taken by the pair occupation measure in \eqref{eq:PairOccupationExplicit}, \eqref{eq:RateFunctionalVariationalCorr} is solved by
\begin{equation}
\label{eq:RateFunctionalLagrangeCorr}
\mathcal{I}_{\ell-1,\ell} \left( \phi_{\ell-1}, \phi_\ell \right) = \sum_{i, j \in V} \pi_{ij} \phi_{\ell-1}(i) u_\ell(i) v_\ell(j) \ln (u_\ell(i) v_\ell(j)) \ ,
\end{equation}
where the vectors $u_\ell$ and $v_\ell$ are solutions of \eqref{eq:EqStochasticityCorr}--\eqref{eq:NormalisationSigma}.

Inserting \eqref{eq:RateFunctionalLagrangeCorr} into \eqref{eq:RateFunctionalCorr} provides the rate at which fluctuations decay as the number of particles $\mathcal{N}$ increases, away from the typical trajectory $(\bar{\rho}_0, \bar{\rho}_1, \ldots, \bar{\rho}_n)$ that solves  \eqref{eq:KineticCorr}. It can be shown that this typical trajectory is a zero of the rate function $\mathcal{I}_{[0,n]}$ in \eqref{eq:RateFunctionalCorr}. To demonstrate this, we proceed step-by-step: given $\bar{\phi}_{\ell-1}$, the zero of \eqref{eq:RateFunctionalLagrangeCorr} is attained at $\phi_{\ell} = \bar{\rho}_\ell$, which solves \eqref{eq:KineticCorr}. It is clear that a zero of \eqref{eq:RateFunctionalLagrangeCorr} is obtained by setting $u(i) = v^{-1}(j) = \text{constant}$. Under this condition, from \eqref{eq:EqForwardChapmanKolmogorovCorr}, we immediately get $\sigma_\ell(j) = \sum_{i \in V} \pi_{ij} \bar{\rho}_{\ell-1}$. Substituting this expression into \eqref{eq:ConstraintRho3Corr}, we recover \eqref{eq:KineticCorr} for all $j \neq s$, while for $j=s$, the solution if fixed by the normalisation condition. Since the underlying stochastic dynamics is ergodic, we argue that this zero is unique, meaning the typical trajectory is the only one that minimises the rate function $\mathcal{I}_{[0,n]}$. 


Solving these equations not only provides the likelihood of observing an occupation $\phi_\ell$ at time $\ell$, given the prior occupation $\phi_{\ell-1}$ at time $\ell-1$, as captured by the rate function \eqref{eq:RateFunctionalLagrangeCorr}, but also reveals the \textit{mechanism} that produces such a fluctuation in a typical manner. From \eqref{eq:EqStochasticityCorr} we identify the transition matrix $\tilde{\Pi}_\ell$ with components 
\begin{equation}
    \label{eq:DrivenProcess}
    \left( \tilde{\pi}_\ell \right)_{ij} = u_\ell(i) \pi_{ij} v_\ell(j) \, ,
\end{equation}
which characterises the dynamics of an effective stochastic process. This process drives $\phi_{\ell-1}(i)$ to $\phi_\ell(i)/\Omega_\ell(i)$, representing the occupation measure before resetting occurs. This relationship is guaranteed by \eqref{eq:EqForwardChapmanKolmogorovCorr}, which expresses the forward Chapman--Kolmogorov equation for this new process. In constrast, Eq.\ \eqref{eq:ConstraintRho3Corr} enforces the condition that the fraction of walkers stepping into vertex $j$ (i.e., $\sigma_\ell(j)$) and surviving at $j$ must match the occupation $\phi_\ell(j)$.

It is important to note that the algebraic structure of \eqref{eq:ConstraintRho3Corr} may impose a constraint either on the occupation measure $\phi_\ell$ that can be studied for a given $\beta$, or on the parameter $\beta$ itself, once the occupation measure $\phi_\ell$ is fixed. 
We will better address this point in the following, before presenting the results.


The case of non-correlated random walks is obtained simply by setting $\beta=0$ in \eqref{eq:EqStochasticityCorr}--\eqref{eq:NormalisationSigma}, Eq.\ \eqref{eq:EqStochasticityCorr} does not change whereas the Chapman--Kolmgorov equation simplifies to
\begin{equation}
    \label{eq:EqForwardChapmanKolmogorov}
    \sum_{j \in V} \phi_{\ell-1}(j) u_\ell(j) \pi_{ji} v_\ell(i) = \frac{\phi_\ell(i)}{\Omega_\ell(i)} \ ,
\end{equation}
where $\Omega_\ell$ in \eqref{eq:OmegaCorr} now reads
\begin{equation}
    \label{eq:Omega}
    \begin{split}
    \Omega_\ell(i) = (1-p) + \delta_{i,s} \frac{p}{\sigma_\ell(i)} \ .
\end{split}
\end{equation}
This highlights that, for $\beta = 0$, a constant fraction $p$ of walkers is reset from any node to the source node, independent of the occupation at those nodes. In general, the $2 N$ equations, comprising \eqref{eq:EqStochasticityCorr} and \eqref{eq:EqForwardChapmanKolmogorov}, must be solved numerically to determine the vectors $u$ and $v$ characterising the effective stochastic process transforming $\phi_{\ell-1}$ into $\phi_\ell$. These solutions also allow for the evaluation of the rate function \eqref{eq:RateFunctionalLagrangeCorr}, which remains valid also for $\beta=0$. 

Regardless of the value of $\beta$, the outlined scheme operates locally in time due to the Markov property, which allows us to express \eqref{eq:RateFunctionalCorr}. This property allows the definition of a sequence $(\tilde{\Pi}_1, \ldots, \tilde{\Pi}_n)$ of locally-in-time protocols, corresponding to the effective stochastic processes given in \eqref{eq:DrivenProcess}, that can generate any occupation measure trajectory $(\phi_0, \phi_1, \ldots, \phi_n)$ as typical.

Before moving to the long-time behaviour of the occupation measure, we discuss the case of a FCG. This will help us develop intuition around the equations \eqref{eq:EqStochasticityCorr}--\eqref{eq:NormalisationSigma}. 

\subsubsection{Illustration via FCG}

In this subsection we showcase the mathematical method in a simple case of a FCG of $N$ vertices. Eqs.\ \eqref{eq:EqStochasticityCorr} and \eqref{eq:EqForwardChapmanKolmogorovCorr} simplifies to
\begin{eqnarray}
    \label{eq:StochasticityFCG}
    u_\ell(i) &=& \frac{N-1}{\sum_{j \neq i} v_\ell(j)} \\
    \label{eq:ForwardChapmanKolmogorovFCG}
    v_\ell(j) &=& \frac{\sigma_\ell(j)}{ (1 - \phi_\ell(j))} H_j^{\phi_{\ell-1}} \, ,
\end{eqnarray}
where 
\begin{equation}
    \label{eq:WeightedHarmonicMean}
            \begin{split}
            H_j^{\phi_{\ell-1}} &= \frac{\sum_{i \neq j} \phi_{\ell-1}(i)}{\sum_{i \neq j} \frac{\phi_{\ell-1}(i)}{x(i)}} \\
            &= \frac{1-\phi_{\ell-1}(j)}{\sum_{i \neq j} \frac{\phi_{\ell-1}(i)}{x(i)}} \, ,
            \end{split} 
\end{equation}
and $x(i) = \sum_{k \neq i} v_\ell(k)$, a weighted harmonic mean. The form \eqref{eq:StochasticityFCG} highlights the local structure of $v_\ell(j)$. 

The form further simplifies if we consider the particular case $\beta=1$ and $p=1$. In such a case, \eqref{eq:ConstraintRho3Corr} admits a solution if and only if 
\begin{equation}
    \label{eq:ConditionFCG}
    \phi_\ell(j) \leq \frac{1}{4} \, .
\end{equation}
This means that selecting $\beta$ puts a constraint on the fluctuations we can study. Assuming \eqref{eq:ConditionFCG} (along with \eqref{eq:ConstraintMinimal}), \eqref{eq:ConstraintRho3Corr} yields the two solutions
\begin{equation}
    \sigma_\ell(j) = \frac{1 \pm \sqrt{1 - 4 \phi_\ell(j)}}{2} \, ,
\end{equation}
both of which are positive and $\leq 1$, and thus, at first glance, both appear acceptable.

However, by introducing an additional necessary constraint that $0 < \sigma_\ell(s) < 1$, which ensures that the total fraction of walkers at the source node $s$ before resetting is positive and less than $1$, we argue that only the solutions with the negative square root are acceptable.

Even in this relatively simple case, we lack a formal mathematical proof for this claim. Our conclusion is instead supported by numerical studies, which consistently indicate the validity of the negative square root solutions.

\subsubsection{Constraints in calculating large deviations and control protocols}
\label{subsubsec:constraints}

Even in the simple case of an FCG, as shown above, an analytical solution to \eqref{eq:EqStochasticityCorr}--\eqref{eq:NormalisationSigma} is out of reach. In general, these equations must be solved numerically.

Given a sequence $(\phi_0, \ldots, \phi_n)$, where all $\phi_\ell$ are positive vectors normalised to $1$ and satisfy \eqref{eq:ConstraintMinimal}, the simplest approach to compute $u_\ell$, $v_\ell$, and $\sigma_\ell$ is to iteratively solve \eqref{eq:EqStochasticityCorr}--\eqref{eq:NormalisationSigma} to convergence. This involves initialising the vectors $u_\ell$ and $v_\ell$ randomly and finding the roots of \eqref{eq:ConstraintRho3Corr}, for each $\ell \in [1, n]$. 

In particular, given a certain $\phi_\ell$, positive real roots of \eqref{eq:ConstraintRho3Corr} are admissible only for $p \geq p_c$ as well as $\beta \geq \beta_c$. To find these critical values it is enough to study \eqref{eq:ConstraintRho3Corr} along with its derivative w.r.t.\ $\sigma_\ell(j)$, i.e.,
\begin{equation}
    \label{eq:DerivativeRho3}
    0 = 1 - p \beta \sigma_\ell^{\beta-1}(j) \, .
\end{equation}
It is easy to see that $\sigma_\ell(j)$ solution of \eqref{eq:DerivativeRho3} is the critical value at which the function 
\begin{equation}
    \label{eq:FunctionConcave}
    g(\sigma_\ell(j)) \coloneqq \sigma_\ell(j) \left( 1 - p \sigma_\ell^\beta(j) \right) - \phi_\ell(j) \, ,
\end{equation}
whose roots we are after, is maximised. Replacing such a solution into \eqref{eq:ConstraintRho3Corr} gives the condition
\begin{equation}
    \label{eq:ConstraintRho3Corrp}
    \left( \frac{1}{p(1+\beta)} \right)^{\frac{1}{\beta}} \left( 1 - \frac{1}{1+\beta} \right) - \phi_\ell(j) = 0 \, ,
\end{equation}
and minimising it for $\beta$ gives $\beta(p) = p^{-1}(1-p)$. By taking $\beta(p)$, substituting it into \eqref{eq:ConstraintRho3Corrp}, and requiring the resulting formula to be non-negative, we obtain the conditions:
\begin{eqnarray}
    \label{eq:pcritical}
    p &\geq& p_c \coloneqq 1 - \phi_\ell(j) \\
    \label{eq:betacritical}
    \beta &\geq& \beta_c \coloneqq \frac{1-p_c}{p_c} \, ,
\end{eqnarray}
which are necessary to ensure positive real roots as solutions of \eqref{eq:ConstraintRho3Corr}. Interestingly, due to the concavity of $g$ in \eqref{eq:FunctionConcave} w.r.t.\ $\sigma_\ell(j)$, and the fact that $\phi_\ell(j)$ only shifts it vertically, it suffices to solve \eqref{eq:ConstraintRho3Corrp} for the largest $\phi_\ell(j)$ in order to calculate $p_c$ and $\beta_c$. However, conditions \eqref{eq:pcritical} and \eqref{eq:betacritical} are necessary but not sufficient to guarantee that the occupation measure before resetting satisfies $0 \leq \sigma_\ell(j) \leq 1$, for all $j \in V$, and is normalised to $1$. Consequently, in the algorithm, an additional step is required to check that the vector $\sigma_\ell$ satisfies these physical constraints.

\subsubsection{Stationary control via long-time limit of sample-path large deviations}

The framework developed so far identifies a sequence of local-time protocols, represented by stochastic matrices $(\tilde{\Pi}_1, \ldots, \tilde{\Pi}_n)$, that drives a system of many random walkers to typically follow a specified trajectory of occupation measures $(\phi_0, \phi_1, \ldots, \phi_n)$ on a general graph $G$. This approach is particularly useful when precise control over the dynamics of random walkers is required.

However, in the resetting literature, the focus is often on stationary properties. Particular attention is given to  the steady state occupation density \cite{evans2020stochastic}, which in the current discrete case corresponds to the \textit{local time} over the graph nodes. This quantity is key to understanding how random walkers distribute their time across the nodes of a graph under resetting dynamics. For this reason, we now shift our focus to stationary properties, focusing on the long-time averages of the occupation measure rather than transient trajectories.

While the mathematics required to solve such problems is generally intricate, we will show how bounding large deviation estimates, rather than computing them exactly, allows for significant simplifications. This approach is customary in application of large deviations, when `contracting' from a higher level to a lower one to focus on sample mean observables, see e.g.,~\cite{Gingrich2016,Gabrielli2020,Pietzonka2023}.

The observable of interest is the time-average of \eqref{eq:OccupationMeasure} and reads
\begin{equation}
\label{eq:OccMeasureTime}
\rho_{n} (i) = \frac{1}{n} \sum_{\ell=0}^n \rho_{\ell}(i) \ ,
\end{equation}
which represents the fraction of time that a fraction of particles spends on each vertex of $G$. We identify this observable as the local time that, instead of referring to a single random walker, captures the behaviour of the entire ensemble of walkers.

Similarly to the earlier analysis, the ultimate goal is to study large deviations of \eqref{eq:OccMeasureTime}. To achieve this, we first scale $\mathcal{N} \rightarrow \infty$ and only then take $n \rightarrow \infty$, as is common in studies of large systems of interacting particles or random walkers on graphs~\cite{Gabrielli2020}. This order of limits allows us to focus on the large-population behaviour of the walkers before considering long-term time averaging.

By taking this approach, we obtain a macroscopic description of the system, where the ensemble of random walkers reaches equilibrium `instantaneously' for any fixed time $n$ as $\mathcal{N} \rightarrow \infty$. This simplifies the description of the system, particularly when individual walkers’ paths are influenced by finite-time correlations due to resetting events (see Eq.\ \eqref{eq:ResetProbability}). In this framework, we prioritise the collective dynamics over the individual temporal trajectories, enabling a clearer understanding of the macroscopic behaviour of the system.

This order of limits yields an LDP for the observable \eqref{eq:OccMeasureTime} of the type
\begin{equation}
\label{eq:LDPEmrOccTime}
\mathbb{P}\left( \rho_{n} = \bar{\phi} \right) = e^{- N n I(\bar{\phi}) + o(n)} \, ,
\end{equation}
with rate function
\begin{equation}
\label{eq:RateFunctionLongTime}
\begin{split}
I(\bar{\phi}) &= \lim_{n \rightarrow \infty} \frac{1}{n} \inf_{\substack{(\phi_1, \cdots, \phi_n) \\ \frac{1}{n} \sum_{\ell=0}^n \phi_\ell  = \bar{\phi} } } \mathcal{I}_{[0,n]} (\phi_0, \phi_1, \cdots, \phi_n) \\
&= \lim_{n \rightarrow \infty} \frac{1}{n} \inf_{\substack{ \phi_\ell \, \forall \ell \in [1,n] \\ \frac{1}{n} \sum_{\ell=0}^n \phi_\ell = \bar{\phi}}} \sum_{\ell=1}^n \mathcal{I}_{\ell-1,\ell} \left( \phi_{\ell-1}, \phi_\ell \right) \ ,
\end{split}
\end{equation}
for normalised $\phi_\ell$s as well as $\bar{\phi}$, which comes from the natural contraction of \eqref{eq:RateFunctionalCorr} requiring that the time average of $\phi_\ell$ equals $\bar{\phi}$.

An exact solution to the variational problem in \eqref{eq:RateFunctionLongTime} is not feasible. [Solving the minimisation problem requires accounting for the implicit derivatives of all the vectors $u_\ell$ in \eqref{eq:EqStochasticityCorr} and $v_\ell$ in \eqref{eq:EqForwardChapmanKolmogorovCorr} with respect to $\phi_\ell$. While it is technically possible, the resulting Euler--Lagrange equations cannot be solved explicitly, and the approach would rely heavily on computationally intense numerics.] However, \eqref{eq:RateFunctionLongTime} can be easily bounded from above by assuming a constant path solution $\phi_\ell = \bar{\phi}$ (which trivially satisfies all constraints), similarly to~\cite{Gabrielli2020}. This gives:
\begin{equation}
\label{eq:BoundRateFunctionLongTime}
I(\bar{\phi}) \leq I^*(\bar{\phi}) \coloneqq \lim_{n \rightarrow \infty} \frac{1}{n} \sum_{\ell=0}^n \mathcal{I}_{\ell-1,\ell} \left( \bar{\phi} , \bar{\phi} \right) \ .
\end{equation}
Because the rate function $\mathcal{I}_{\ell-1,\ell} \left( \bar{\phi}, \bar{\phi} \right)$ does not depend on time in this case, the bound can be simplified by substituting $\bar{\phi}$ into \eqref{eq:RateFunctionalLagrangeCorr}, resulting in:
\begin{equation}
\label{eq:BoundRateFunctionFinal}
I^*(\bar{\phi}) = \sum_{i, j \in V} \pi_{ij} \bar{\phi}(i) u(i) v(j) \ln (u(i) v(j)) \ ,
\end{equation}
where the vectors $u$ and $v$ are solution of the following time-independent system of equations, analogous to \eqref{eq:EqStochasticityCorr}--\eqref{eq:NormalisationSigma} and for simplicity repeated below
\begin{eqnarray}
\label{eq:EqStochasticityCorrStat}
u(i) &=& \frac{1}{\sum_{j \in V} \pi_{ij} v(j)} \\
\label{eq:EqForwardChapmanKolmogorovCorrStat}
v(j) &=& \frac{\sigma_\ell(j)}{\sum_{i \in V} \pi_{ij} \bar{\phi}(i) u(i)} \\
\label{eq:ConstraintRho3CorrStat}
\bar{\phi} (j) &=& \sigma(j) \left( 1-p \sigma^\beta(j) \right) \, \, \, \, \forall j \neq s \\
\label{eq:NormalisationSigmaStat}
\sigma(s) &=& 1 - \sum_{i \neq s} \sigma(i) \, .
\end{eqnarray}
To verify the validity of \eqref{eq:BoundRateFunctionLongTime}, we replace $\bar{\phi} = \bar{\rho}_{\text{inv}}$, where $\bar{\rho}_{\text{inv}}$ is the solution of \eqref{eq:Stationary}, into \eqref{eq:BoundRateFunctionLongTime}--\eqref{eq:NormalisationSigmaStat}. This yields $I^*(\bar{\phi} = \bar{\rho}_{\text{inv}}) = 0$, confirming consistency with the known stationary state.

Similarly to the earlier case, in addition to providing a bound on the likelihood of visiting a fluctuation $\rho_n = \bar{\phi}$ at large times $n$, solving \eqref{eq:EqStochasticityCorrStat}--\eqref{eq:NormalisationSigmaStat} for $u$ and $v$ identifies a stochastic mechanism, characterised by the transition matrix
\begin{equation}
    \label{eq:DrivenProcessStat}
    \tilde{\pi}_{ij} = u(i) \pi_{ij} v(j) \, ,
\end{equation}
which generates $\bar{\phi}$ as a typical (stationary) local time for the system of random walkers on the graph $G$.

Unlike before, since we are bounding the true large deviation function from above, the long-time protocol defined by \eqref{eq:DrivenProcessStat} is not an `optimal' protocol in the large deviation sense. Specifically, its path probability, leading to $\bar{\phi}$ in the long-time limit, is not the closest in terms of relative entropy to the path probability of the original process~\cite{Chetrite2015a}. Nevertheless, for the purposes of this work, this is not a significant limitation. While we cannot fully characterise large deviations, \eqref{eq:DrivenProcessStat} provides a sufficient mechanism for exploring all scenarios of interest.

Before proceeding to discuss results, we note that the constraints highlighted in Section \ref{subsubsec:constraints}, will remain equally significant here in solving \eqref{eq:EqStochasticityCorrStat}--\eqref{eq:NormalisationSigmaStat}. These constraints continue to ensure the physical consistency of the occupation measures.

\section{Illustrative examples and discussion}
\label{sec:Examples}

In the following, we illustrate the framework developed above with examples involving a FCG and one heterogeneous graph structure.

Our primary focus will be on characterising the likelihood and stationary control processes that lead to stationary local times, rather than entire occupation measure trajectories. This emphasis is motivated by the fact that stationary local times are particularly relevant in resetting dynamics, where the long-term distribution of time spent on each vertex captures the essential macroscopic behaviour of the system. Unlike transient trajectories, stationary properties provide insight into the steady-state organisation of walkers, are less sensitive to initial conditions, and may therefore be more robust and meaningful for many applications.

That said, it is important to emphasise that the framework and theory developed here are far more general. They allow for control over entire occupation measure trajectories, providing a detailed and versatile approach to studying random walker dynamics. The focus on stationary properties in this last section reflects the relevance and tractability of these properties for practical applications, but the general framework justifies the broader scope and detail devoted to its development earlier in the paper.

\subsection{Fully connected graphs}

\begin{figure}
    \centering
    \includegraphics[width=0.7\linewidth]{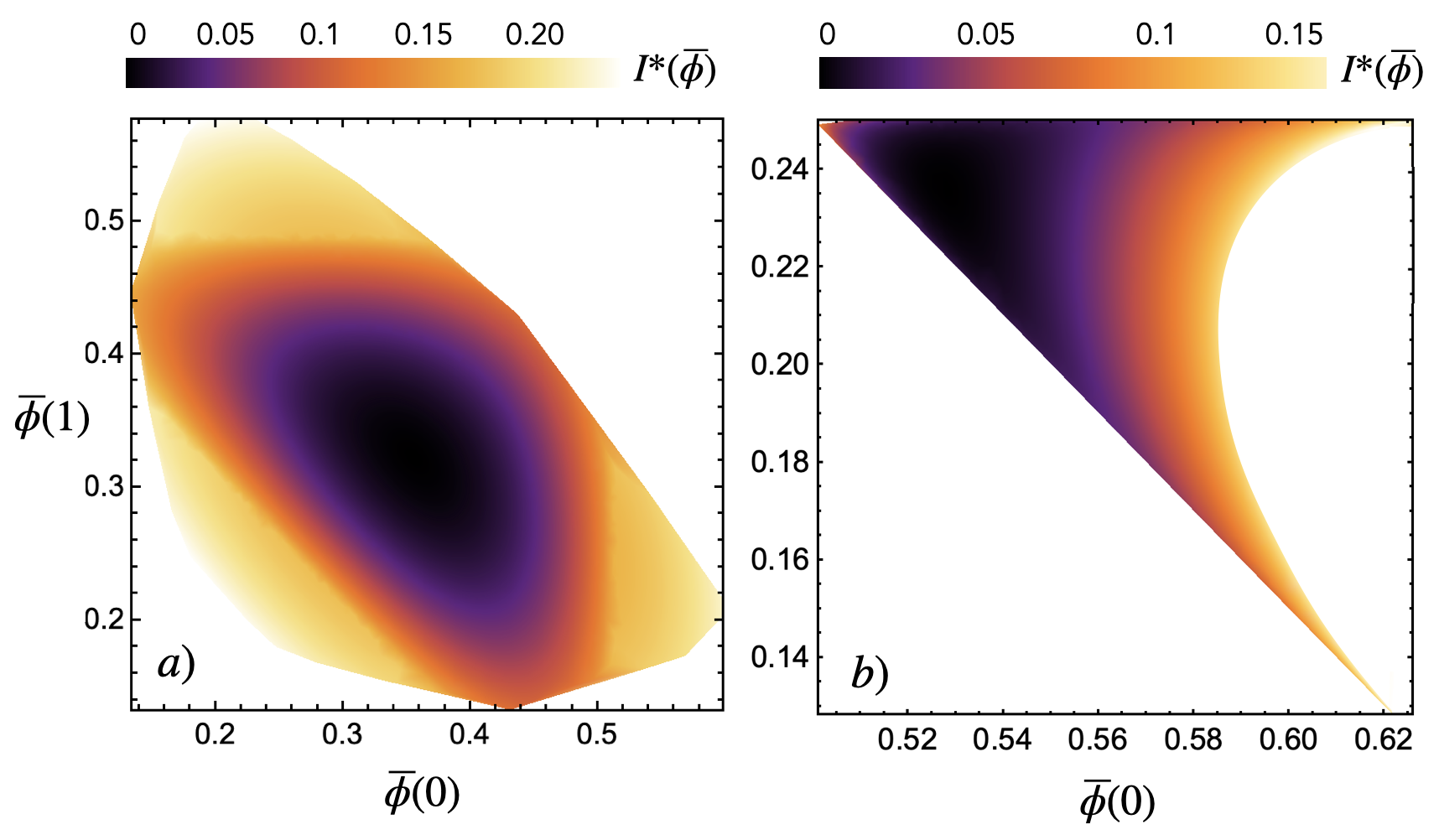}
    \caption{Contour plot of rate function $I^*(\bar{\phi})$ for $10^4$ random realisations of $\bar{\phi}$ on a FCG of $3$ vertices, labelled $0,1,2$ for a reset protocol $p=0.05$ and $\beta=0$ in a) and $p=1$ and $\beta=0.5$ in b).}
    \label{fig:3state}
\end{figure}

The rate function \eqref{eq:BoundRateFunctionFinal} $I^*$ takes value in $[0,1]^N$ and maps it in $\mathbb{R}^+$. As such, for large $N$, it is a complicated object to visualise. We restrict to $N=3$, random sample $\bar{\phi} \in [0,1]^3$ and calculate $I^*$ by plugging in $u$ and $v$ solution of \eqref{eq:EqStochasticityCorrStat}--\eqref{eq:NormalisationSigmaStat} considering two cases, $\beta=0$ and $p=0.05$, and $\beta=1.$ and $p=1$ (source node is labeled as $0$). Because of the normalisation condition $\sum_{i \in V} \bar{\phi}(i) = 1$, the possible values that $\bar{\phi}$ can take live on a 2D plane of $[0,1]^3$. Such a constraint effectively restrict the domain of the rate function $I^*$ to $[0,1]^2$ for visualisation purposes. We plot the rate functions in Fig.\ \ref{fig:3state}. [We remark that not even in such a simple scenario of a complete graph of $3$ vertices it is possible to solve \eqref{eq:EqStochasticityCorrStat}--\eqref{eq:NormalisationSigmaStat} analytically.]


Both rate functions are convex and non-negative with single minima. These minima are also zeros of the rate functions and therefore characterise the typical values of the local-time random variable $\rho_n$ in \eqref{eq:OccMeasureTime} that a system composed by a large number of unbiased random walkers, with the particular reset scheme considered, reach at long times. The typical values can be calculated analytically for the case $\beta=0$, via \eqref{eq:StationaryFully}, and numerically for $\beta=1$, using \eqref{eq:SysStatFully}. These values are $\approx (0.379, 0.311, 0.311)$ and $\approx (0.528, 0.236, 0.236)$, respectively, and evidently sit at the minimum of the rate function.

Atypical configurations are all those for $I^* > 0$. The greater the value of $I^*$, the more atypical a configuration is. White regions on the plot represent physically unrealisable states due to the constraints discussed in Section~\ref{subsubsec:constraints}. Specifically, for $\beta = 1$, it is nearly impossible to observe states where the population at the source node is less than typical. Such a configuration is obtained in the most likely way by resetting a balanced distribution of walkers over nodes $1$ and $2$, as the population returning to node $0$ scales quadratically for $\beta = 1$, i.e., $\bar{\phi}(1)^2 + \bar{\phi}(2)^2$. To achieve a minimal population on the source node, $\bar{\phi}(1)$ and $\bar{\phi}(2)$ must also be minimal. Conversely, asking the population at the source node to be larger than typical requires maximizing the resetting contribution from the other two nodes. This scenario is achieved by creating an unbalanced population distribution over nodes $1$ and $2$, as evidenced by the curvilinear shape in Fig.\ \ref{fig:3state}b.

This behaviour highlights the intricate interplay between fluctuation and reset mechanisms. Furthermore, such a simple study proves that the framework developed above works as well as that the underlying dynamics can be very rich even for simple systems. We now move to larger graphs and aim to answer the interesting question on whether given a flat $\bar{\phi}$, viz.\ the local-time is the same over all vertices but the source one, there exists a density-dependent reset scheme characterised by a particular value of $\beta$ that allows to realise such a \textit{flat state} as typical. This is a first step towards understanding whether it is possible to use the framework introduced above to avoid large heterogeneity in the local time, therefore preventing overcrowdedness in sub-portions of the graph.

\begin{figure}
    \centering
    \includegraphics[width=0.7\linewidth]{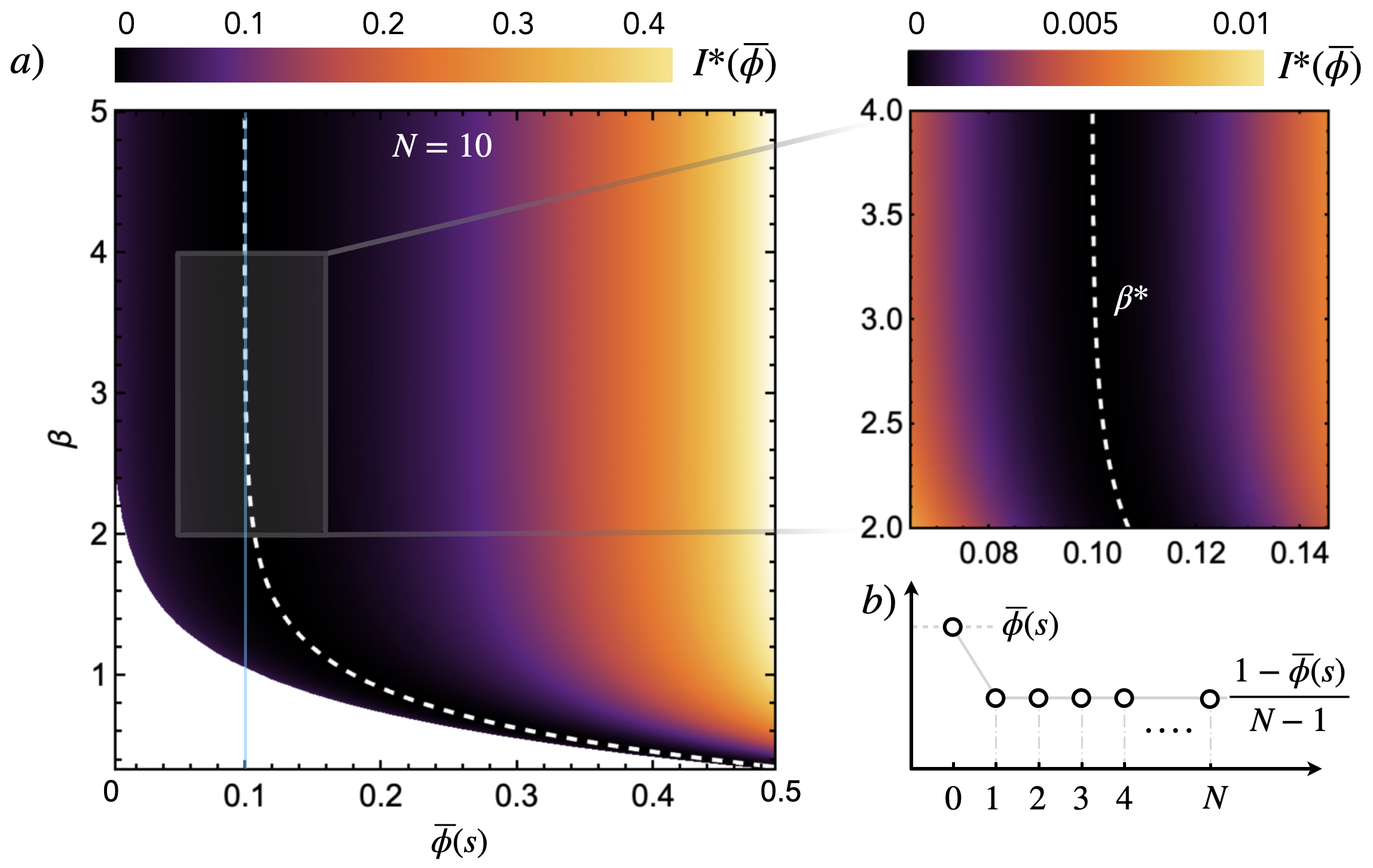}
    \caption{Rate function $I^*$ plotted as a function of $\beta$ and $\bar{\phi}(s) \equiv c$ for a FCG of $N=10$ nodes for a flat-state occupation such as the one in panel b). The white dashed line marks the optimal protocol scheme $\beta^*(c)$ which corresponds to the minimum of the rate function. The solid blue line at $1/N$ marks the typical occupation of the source node in the case of no reset. }
    \label{fig:IVSBeta_FCG}
\end{figure}

We consider $N=10$ and impose 
\begin{equation}
    \label{eq:FlatState}
    \bar{\phi}(i) = \delta_{i,s} c + (1-\delta_{i,s}) (1 - (N-1)c) \, , 
\end{equation}
for constants $c$ (see Fig.\ \ref{fig:IVSBeta_FCG}b) and calculate, for each $c$ and $\beta$ over a fine mash of $(0,10]$,  $u$ and $v$ from \eqref{eq:EqStochasticityCorrStat}--\eqref{eq:NormalisationSigmaStat} and $I^*$ from \eqref{eq:BoundRateFunctionFinal}. It is hard to guess a-priori the form of such a function, but we expect, given the fully connected structure, for each $c$ to find a zero at a value $\beta^*(c)$ which characterises a density-dependent reset scheme that would make the respective $\bar{\phi}$ typical. We plot $I^*$ as a function of $\beta$ and $\bar{\phi}(s) \equiv c$ in Fig.\ \ref{fig:IVSBeta_FCG}a. Along with it, we plot $\beta^*(c)$, which can be calculated numerically inverting \eqref{eq:Stationary}, as a white dashed line lying at the bottom and minimum of $I^*$, and $1/N$ as a blue line marking the asymptotic occupation of the source node for $\beta \rightarrow \infty$, i.e., no reset. 

Evidently, as expected, for each $\bar{\phi}(s)$ there exists an optimal density-dependent reset scheme that makes the respective $\bar{\phi}$ typical. Furthermore, at fixed $\bar{\phi}(s)$, $I^*$ is not monotonic. As $\beta \to \infty$, it asymptotically approaches a constant value, indicating that excessively large $\beta$ values do not significantly impact the system. We also note that the optimal values of $\beta^*$ all lie above $1/N$. In other words, it is not possible to find a $\beta$ such that the typical occupation has a local time less than $1/N$ on the source node. [The white region in Fig.\ \ref{fig:IVSBeta_FCG}a marks the non-physical combinations of $\bar{\phi}$ and $\beta$ for which the rate function does not exist.]

We now move to the case of heterogeneous random graphs, where typically one may record overcrowdedness on certain portions of the graph and we address the question on how a density-dependent reset scheme can influence the appearance of overcrowded states.

\subsection{Heterogeneous graphs}

We focus on the exemplary class of heterogeneous random graphs introduced in \ref{sec:BA}: BA random graphs.

We consider the graph depicted in Fig.\ \ref{fig:BA_typical}a with $N=20$ vertices. As discussed previously, by simulating a system composed by a large number of unbiased random walkers, one records the local times plotted in Fig.\ \ref{fig:BA_typical}b. Naturally, due to the occurrence of vertices that are more connected than others---hubs are a common feature of BA graphs---the ensemble of random walkers tend to spend more time on these vertices and less on others.

In this context, the likelihood of a flat state, such as \eqref{eq:FlatState}, becoming more likely to appear is of interest. By adjusting the reset parameter $\beta$, it is possible to influence the likelihood of achieving a flat state.

\begin{figure}[h!]
    \centering
    \includegraphics[width=0.7\linewidth]{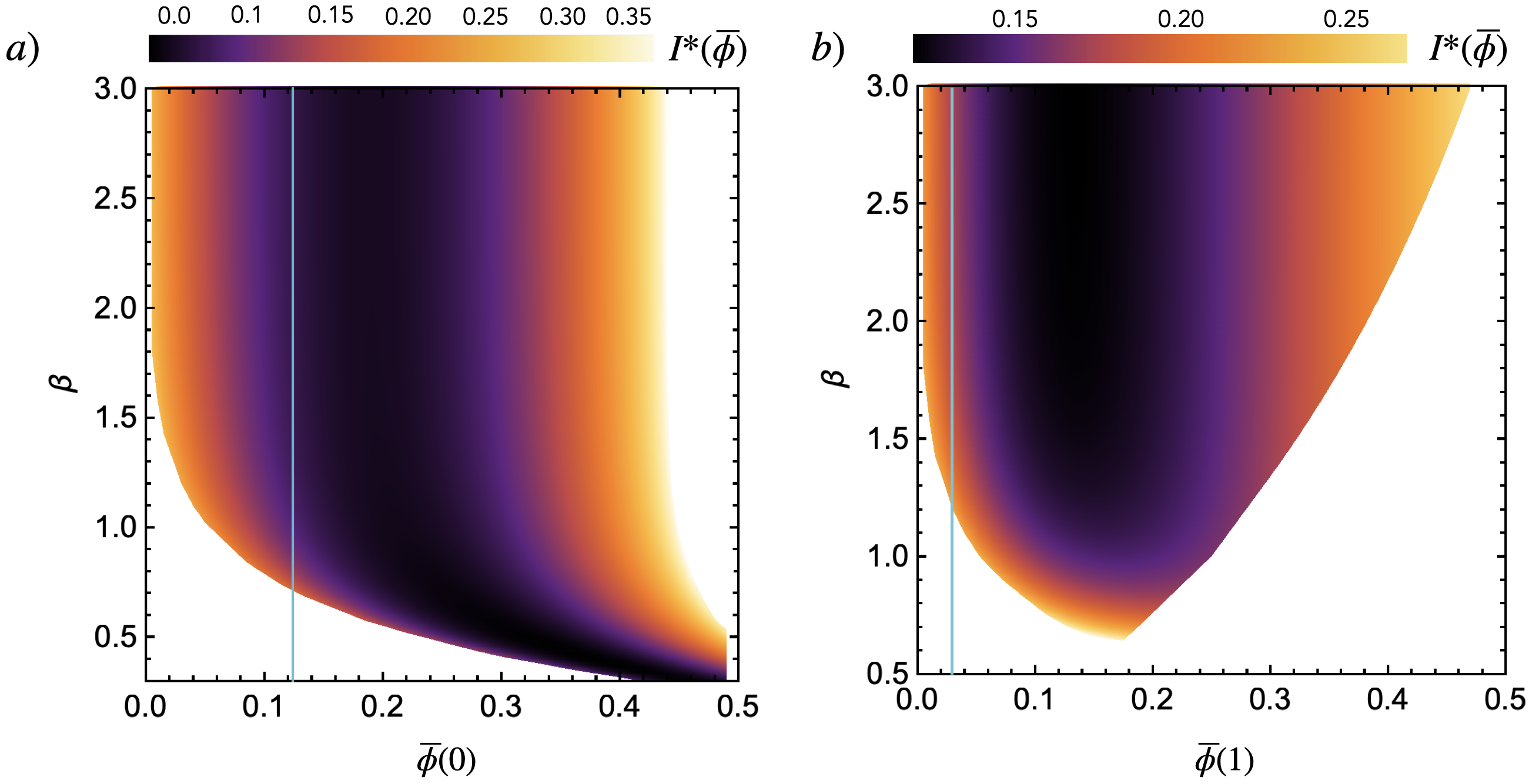}
    \caption{Rate function $I^*$ plotted as a function of $\beta$ and $\bar{\phi}(s) \equiv c$ for a BA graph of $N=20$ nodes. The flat-state occupation $\bar{\phi}$ is given in \eqref{eq:FlatState}, with two scenarios considered: panel (a) corresponds to the source node $s=0$, the hub of the graph, while panel (b) corresponds to $s=1$, a node with few connections. The solid blue line indicates the typical occupation of the source node in the absence of reset.}
    \label{fig:IVSBeta_BA}
\end{figure}

Due to the heterogeneity of the graph, the statistical properties of vertices differ, and the choice of source vertex plays a role. To explore this, two flat state scenarios are considered: (i) where the source vertex is the hub of the graph, labeled as $s=0$ in Fig.\ \ref{fig:BA_typical}a, and (ii) where the source vertex is one of the least connected nodes of the graph, e.g., node $s=1$. For each scenario, $\bar{\phi}$ is taken as in \eqref{eq:FlatState} with different values of $c$. The rate functions $I^*$ at the flat state are then calculated as a function of $\beta$, and the results are shown in Fig.\ \ref{fig:IVSBeta_BA}.

In general, the rate functions no longer exhibit a zero as a function of $\beta$. This outcome directly results from the heterogeneous connections in the graph, meaning that a $\beta$-dependent reset scheme can no longer generate a flat state as a typical state. However, it is still possible to identify a reset scheme that minimises the rate function, thereby maximising the probability of a flat state and making it exponentially more likely to occur as a fluctuation under such a reset scheme (the darker the colour in the figure, the more likely the state to appear). [As before, the white region in Fig.\ \ref{fig:IVSBeta_BA} indicates the non-physical combinations of $\bar{\phi}$ and $\beta$ for which the rate function $I^*$ does not exist.]

Evidently, from both plots, increasing $\beta$ excessively has little impact. This is because larger values of $\beta$ make it harder to reset from nodes with moderate occupation (which comprise the majority in the typical scenario), resulting in minimal changes to the rate function.

When the source node is the hub, Fig.\ \ref{fig:IVSBeta_BA}a, the lower the occupation on it, the harder it is to achieve such a scenario. Consequently, larger $\beta$ values are required for reset schemes to make this scenario possible, at least in the fluctuations. Conversely, when the hub is heavily occupied, lower $\beta$ values suffice to achieve a flat state, as stronger resets are necessary even on moderately occupied nodes to redistribute walkers and achieve high occupation on the source node.

For the case where the source node is one of the least connected nodes, Fig.\ \ref{fig:IVSBeta_BA}b, the situation appears partially opposite. Achieving a flat state with high occupation on a poorly connected source node becomes increasingly less physically realisable. In this case, the reset must be weaker for nodes with low occupation but relatively stronger for nodes with high occupation to compensate. This arises because walkers reset to a poorly connected source node are not well redistributed across the network in the immediately subsequent time steps, leading to persistent overcrowding near the source node and making a homogeneous state difficult to achieve.

On the other hand, achieving a flat state with low occupation on a poorly connected source node is easier (more physically realisable) as a fluctuation. The lower the occupation required, the higher the $\beta$ value needed to reduce frequent resetting, which would otherwise increase the source node's local time.

\section{Conclusion and Outlook} \label{sec:Conclusion}

In this paper, we developed a novel framework for designing density-dependent resetting protocols aimed at achieving specific occupation measure configurations over networks. Using concepts from stochastic resetting and large deviation theory, we explored how the resetting scheme, governed by a power-law dependence on the local density of random walkers, influences the likelihood of achieving desired configurations, such as homogeneous (or flat, as named in the paper) distributions. These are characterised by a constant occupation measure over the whole network, except at the source node, which acts as a warehouse stocking walkers and facilitating their redistribution over the network. 

The framework was applied to a fully connected and an exemplary heterogeneous network, illustrating the interplay between network structure and resetting protocols. We showed that density-dependent resetting schemes could significantly enhance the probability of otherwise rare configurations. We have seen that achieving flat configurations on networks with hubs or poorly connected nodes is generally harder.

Our study opens several avenues for future research:

\begin{itemize}
    \item Future work could investigate the typical configurations emerging from density-dependent resetting in simpler systems, such as one-dimensional diffusion processes. These setups might offer other analytical insights and help elucidate the mechanisms underlying the observed effects in more complex networks.

    \item While we focused on homogeneous distributions, the framework can readily be adapted to target other configurations. For example, one might design resetting protocols to maximise population accumulation on specific nodes, which could be relevant for tasks requiring node-specific resource concentration.

    \item Our resetting mechanism is deterministic given a node’s population, but including additional stochastic effects in the resetting rule could add another layer of control. For instance, allowing reset probabilities to themselves fluctuate could provide different pathways to optimise the likelihood of rare events as combining random resetting with random exploration might enhance the system's ability to achieve desired configurations.

    \item A key open question is the systematic identification of the optimal $\beta^*$ for a given graph $G$ and desired configuration $(\phi_0, \cdots, \phi_n)$ or $\bar{\phi}$. At the moment, we do not have an analytical tool to determine the optimal $\beta^*$, which opens up to further investigation.

    \item While our study addressed global local-time distributions, future work could focus on single-vertex local times. Understanding how resetting impacts the local-time statistics of individual vertices could provide deeper insights into the interplay between network structure and resetting dynamics.

    \item The framework could be extended to study search strategies in random walk dynamics, where trajectories associated with low occupation densities over the graph are reset less frequently, potentially improving the likelihood of finding distant targets. Such optimisation could be critical in applications like resource allocation or information retrieval on networks.

\end{itemize}

In conclusion, our study establishes a framework for leveraging density-dependent resetting protocols to control population distributions over networks. By combining theoretical insights with practical examples, we lay the groundwork for future studies that could refine these methods and explore their applications in various networked systems. 

\acknowledgements
FC thanks Gr\'{e}gory Schehr for insightful discussions during the 2024 Les Houches Summer School `Theory of Large Deviations and Applications'. FC acknowledges funding through EPSRC Grant No.\ EP/V031201/1 and KSO support from the Alexander von Humboldt foundation.

\appendix

\section{Derivation explicit rate function \label{Appendix:RateFunctionHigher}}

The calculation of \eqref{eq:RateFunctionHigher} uses G\"{a}rtner--Ellis theorem~\cite{Touchette2009,Dembo2010,Touchette2018}, which states that if the Scaled Cumulant Generating Function (SCGF), defined as 
\begin{equation}
    \label{eq:SCGFDef}
    \Psi_{\phi_{\ell-1}}(\zeta) = \lim_{\mathcal{N} \rightarrow \infty} \frac{1}{\mathcal{N}} \ln \mathbb{E}_{\phi_{\ell-1}} \left[ e^{\mathcal{N} \sum_{i,j \in V} \zeta(i,j) L_\ell(i,j)} \right] \, ,
\end{equation}
exists and is differentiable, then the rate function associated with the random variable pair occupation measure can be calculated via the Legendre–Fenchel transform of $\Psi_{\phi_{\ell-1}}(\zeta)$. Here, $\mathbb{E}_{\phi_{\ell-1}}$ refers to the expectation conditioned on $\phi_{\ell-1}$.

The SCGF can be calculated as follows:
\begin{equation}
\label{eq:SCGFCalculation}
\begin{split}
\Psi_{\phi_{\ell-1}}(\zeta) 
&= \lim_{\mathcal{N} \rightarrow \infty} \frac{1}{\mathcal{N}} \ln \mathbb{E}_{\phi_{\ell-1}} \left[ e^{\sum_{k=1}^\mathcal{N} \zeta \left( X^{(k)}_{\ell-1},\tilde{X}^{(k)}_{\ell} \right) } \right] \\
&= \lim_{\mathcal{N} \rightarrow \infty} \frac{1}{\mathcal{N}} \sum_{k=1}^N \ln \sum_{j \in V} \pi_{X^{(k)}_{\ell-1},j} e^{\zeta \left( X^{(k)}_{\ell-1},j \right) } \\
&= \sum_{i \in V} \phi_{\ell-1}(i) \ln \sum_{j \in V} \pi_{i,j} e^{\zeta(i,j)} \, .
\end{split}
\end{equation}
In the above, we use the definition of the pair occupation measure $L_\ell(i,j) = \frac{1}{\mathcal{N}} \sum_{k=1}^N \delta_{X^{(k)}_{\ell-1},i} \delta_{\tilde{X}^{(k)}_{\ell},j}$, (where $\tilde{\cdot}$ refers to the pre-reset state), the fact that particles move independently, and the large-$\mathcal{N}$ limit conditioned on the density $\phi_{\ell-1}$. 

From the SCGF in the last line of \eqref{eq:SCGFCalculation}, we can calculate $\mathcal{H}(L_\ell|\phi_{\ell-1})$ using the Legendre–Fenchel transform:
\begin{equation}
\label{eq:LegendreTransform}
\mathcal{H}(L_\ell|\phi_{\ell-1}) = \sup_{\zeta} \left( \sum_{i,j \in V} \zeta(i,j) L_\ell(i,j) - \Psi_{\phi_{\ell-1}}(\zeta(i,j)) \right) \ .
\end{equation}
To solve it, we set the derivative to zero:
\begin{equation}
\label{eq:Variation0}
\frac{\partial \left( \sum_{i',j' \in V} \zeta(i',j') L_\ell(i',j') - \Psi_{\phi_{\ell-1}}(\zeta(i',j')) \right)}{\partial \zeta(i,j)} = 0 \ ,
\end{equation}
which yields the Euler--Lagrange equation
\begin{equation}
\label{eq:EulerLagrange}
L_\ell(i,j) - \frac{\phi_{\ell-1}(i) \pi_{i,j} e^{\zeta(i,j)}}{\sum_{j \in V} \pi_{i,j} e^{\zeta(i,j)}} = 0 \ .
\end{equation}
Summing over $j$, we recover the condition $\phi_{\ell-1}(i) = \sum_{j \in V} L_\ell(i,j)$.

Finally, formally solving for $\zeta(i,j)$ from \eqref{eq:EulerLagrange} and substituting it back into \eqref{eq:LegendreTransform}, we recover \eqref{eq:RateFunctionHigher}.

\newpage

\section{Large-Deviation Algorithm Pseudocode}
\label{Appendix:Algorithm}

\begin{algorithm}[H]
\caption{Large-Deviation Algorithm}
\begin{algorithmic}[1] 
\Function{Normalised\_Vector}{vector}
    \State Normalises a given vector to \( 1 \).
\EndFunction
\Function{Rnd\_Gen}{}
    \State Generates a random number.
\EndFunction
\Function{Beta\_Critical}{$\beta$, index, $\phi_\ell$, $p$}
    \State Calculates \( \beta_c \).
\EndFunction
\Function{Sigma\_Fun}{$\sigma_\ell$, index, $\beta$, $\phi_\ell$, $p$}
    \State Defines the main equation for \( \sigma_\ell \).
\EndFunction
\Function{U\_Fun}{$v_\ell$, index, N}
    \State Computes \( u_\ell[\text{index}] \) at iteration \( m \).
\EndFunction
\Function{V\_Fun}{$\phi_{\ell-1}$, $u_\ell$, $\sigma_\ell$, index, N}
    \State Computes \( v_\ell[\text{index}] \) at iteration \( m \).
\EndFunction

\Statex 

\State \textbf{Initialise Parameters:}
\State Set graph \( G \), size \( N \), source node \( s \), reset parameters \( p \) and \( \beta \), and given \( \phi_{\ell-1} \).
\State Initialise \( \phi_\ell \) as a normalised random vector or chosen normalised vector.

\If{\( \phi_\ell[0] < p \left(\frac{1}{N-1}\right)^{\beta} \)}
    \State Adjust \( \phi_\ell[0] \) and iterate through \( \phi_\ell \).
\EndIf

\State Compute \( p_c = 1 - \phi_\ell[\texttt{index\_max}] \).
\If{\( p \leq p_c \)} \State \textbf{Abort}. \EndIf

\State Solve for \( \beta_c \) using \texttt{Beta\_Critical}.
\If{\( \beta \leq \beta_c \)} \State \textbf{Abort}. \EndIf

\For{each index \( \in V \)}
    \State Compute \( \sigma_\ell[\text{index}] \).
\EndFor

\Statex 

\While{error \( > \) tolerance}
    \State Update \( u_\ell \) and \( v_\ell \).
    \State Calculate error.
\EndWhile

\State Output final results: \( \sigma_\ell \), \( u_\ell \), \( v_\ell \).

\end{algorithmic}
\end{algorithm}

\newpage

\section*{References}
\bibliography{mybib.bib}

\end{document}